\documentclass[12pt]{article}
\usepackage{amsmath}
\usepackage{epsfig}
\begin{document}
\title{Dependence of nucleon mass on parameters of instanton medium
}
\author{E. G. Drukarev, E. L. Kryshen, M. A. Matveev, \\
M. G. Ryskin, V. A. Sadovnikova\\
{\em National Research Center "Kurchatov Institute"}\\
{\em B. P. Konstantinov Petersburg Nuclear Physics Institute}\\
{\em Gatchina, St. Petersburg}}
\date{}
\maketitle

\begin{abstract}
We trace the dependence of the nucleon mass and of the effective size of the nucleon on the characteristics of the instanton medium, which are the average instanton size $\rho$ and the distance between the instantons in instanton medium $R$.
We consider several scenario for variation of the values of $\rho$ and $R$.
\end{abstract}

\section{Introduction}

One of the main achievements of the QCD sum rules (SR) approach \cite{1}, \cite{2}, \cite{3} is that it enabled to express the nucleon mass in terms of the QCD quark condensates. The nucleon mass was found to depend mainly on the value of the scalar quark condensate  $\langle0|\bar q(0)q(0)|0\rangle$. The approach is based on the dispersion relations
for the function $\Pi(q)$ describing the propagation of the system with the four-momentum $q$ and the quantum numbers of the hadron. In the case of proton there are two functions $\Pi^q$ and $\Pi^I$, corresponding to the two structures in the propagator of a fermion. The dispersion relations are considered at $q^2\to-\infty$. This makes possible to carry out expansion of  the functions $\Pi^i(q)$ in powers of $1/q^2$. The coefficients of the expansion are the QCD condensates. This is known as the Operator Product Expansion (OPE). It is important that in expectation values the QCD operators
are taken in the same space-time point.

However, the structure of the QCD vacuum is more complicated \cite{3}, \cite{4}. There are strong gluon fields localized on the distances $\rho$  of the order $0.3~$fm which are known as instantons. The instantons produce the quark-antiquark pairs which are present in the QCD vacuum with a nonzero density. At the large distances $r \gg \rho$ one deals with the
averaged density $\langle 0|\bar q(0)q(0)|0\rangle$ in which both quarks are at the same point. However at smaller distances $r \leq
\rho$ one should take into account that the quark operators act in different points. In other words one should include the instantons into the SR.

The contribution of the instantons into the SR were considered  in \cite{5}, \cite{6}, \cite{7}, \cite{8}. Actually the authors of \cite{5}-\cite{7} were interested in description of the quark correlations in the polarization operator due to interaction with the same instanton.  In other words the authors considered the instanton correction to the Green function of {\em two} quarks but neglect the instanton effect in the propagation of one quark where just the "mean field" scalar condensate
$\langle0|\bar q(0)q(0)|0\rangle$ was included. In \cite{8} the authors calculated the polarization operator of the nucleon current
basing on description of the light quark in instanton medium developed in \cite{9}, \cite{10}. In papers \cite{9},\cite{10} the QCD vacuum contains a correlated system of instantons and antiinstantons of the size $\rho \approx 0.3~$fm separated by the distances $R \approx 1$ fm.

In this approach the propagator of a light ($u$ or $d$) quark is described by the Green function, written in Euclidean space as
\begin{equation}
S(p)=\frac{\hat p+im(p)}{p^2+m^2(p)}=S_p+iS_I; \quad S_p=\frac{\hat p}{p^2+m^2(p)}; \quad S_I=\frac{m(p)}{p^2+m^2(p)}.
\label{1}
\end{equation}
The quark mass $m(p)$ depends the average instanton size $\rho$. It is also proportional to the density of the instantons
$N/V=1/R^4$, where $R$ is the average distance between the instantons. The calculations in \cite{8} were carried out at fixed values
\begin{equation}
\rho=\rho_0=0.33~{\rm fm}; \quad R=R_0=1~{\rm fm},
\label{4}
\end{equation}
i.e. $\rho \approx (600$ GeV$)^{-1}=1.69$ GeV$^{-1}$, $R \approx (200$ GeV$)^{-1}=5.07$ GeV$^{-1}$, providing reasonable value for the nucleon mass.

However, there is certain distribution in $\rho$ around the value $\rho_0$. Also, the value of $R$
is connected with the value of the gluon condensates which is assumed to be totaly produced by the instantons
\begin{equation}
\frac{1}{32\pi^2}\langle 0|G^{a\mu \nu}G^a_{\mu \nu}|0\rangle=\frac{N}{V} \equiv\frac{1}{R^4}.
\label{3aa}
\end{equation}
The value of this condensate is also known with certain errors. Thus one needs to analyze the dependence of the nucleon mass on possible variation of these parameters. It is also important to study the behavior of the nucleon residue $\lambda^2$ during these variations. Assuming that the three valence quarks of the nucleon can be described in terms of quantum mechanics, the value of $\lambda^2$ can be identified with the square of the wave function $\Psi$ of the three quarks at the origin. If there is one scale of distances $r_N$, it can be treated as the size of the nucleon  one can put $|\Psi(0)|^2 \sim r_N^{-6}$. Thus the change of the value of $\lambda^2$ by the value of $\delta \lambda^2$ leads to the change of the size of the nucleon
\begin{equation}
\frac{\delta r_N}{r_N}=-\frac{1}{6}\frac {\delta \lambda^2}{\lambda ^2}.
\label{3bb}
\end{equation}
Also, variation of the effective threshold $W^2$ can be treated as variation of masses of the higher states with the quantum numbers of the baryons.

We carry out our analysis for two cases. First we neglect the correlations between the quarks in polarization operator,
describing each of them by the propagator (\ref{1}). In the second case we include the influence of the gluon condensate on the two quarks simultaneously. In \cite{9},\cite{10} all the gluon condensate is assumed to be originated by instantons. Its action on each quark separately is included in the propagator (\ref{1}). However, there is also a contribution corresponding to the gluons interacting  with different quarks of the polarization operator \cite{11}.
Note that inclusion of the averaged gluon condensate $\langle 0|G^{a\mu \nu}G^a_{\mu \nu}|0\rangle $ is not selfconsistent in this approach. The gluon field is also inhomogeneous and should be described by an explicit instanton configuration. The described way of inclusion of correlations beyond the independent particle approach overestimates its effect.

As a first step we can vary the values of $\rho$ and $R$ independently. However, these values are related by the condition of equilibrium of the instanton medium and should be varied simultaneously. We demonstrate that variation of the parameters can change the value of the nucleon mass by about $50~$MeV.

In Sec. 2 we present the dependence of the SR equations on $\rho$ and $R$. In Sec.~3 we vary the parameters in independent way. In Sec.~4 we vary them in a selfconsistent way. We summarize in Sec.~5.

\section{Sum rules}

The Borel transformed QCD SR in the instanton medium can be written as \cite{8}
\begin{equation}
\tilde A'_0(M^2, W^2)+ \tilde A'_4(M^2, W^2)+\tilde A'_6(M^2) =\lambda^2e^{\frac{m^2}{M^2}} ,
 \label{20}
\end{equation}
and
\begin{equation}
\tilde B'_3(M^2, W^2)+
\tilde B'_6(M^2)=m\lambda^2e^{\frac{m^2}{M^2}}.
\label{21}
\end{equation}
Here Eq.~(\ref{20}) corresponds to the $\hat q$ structure of the polarization operator, while Eq.~(\ref{21})  is for the $I$ structure. The lower indices denote the dimensions of the condensates in the OPE version of the SR.
The contributions to Eq.~(\ref{20}) contain even number $n$  of propagators $S_I$ ($n=0,2$).
The terms with odd $n=1,3$ contribute to Eq.~(\ref{21}).

In the term $\tilde A'_0(M^2, W^2)$ the three quarks are described by the propagator $S_q$ - see Eq.~(\ref{1}).
\begin{equation}
\tilde A_0'(M^2,W^2)=M^6F_0(M^2, W^2)/L(M^2),
\label{22}
\end{equation}
$$F_0(M^2)=E_2(\gamma)-\frac{36I_3E_0( \gamma)}{M^4}+\frac{24I_5}{M^6}-\frac{3(I_7-16I_3^2)}{M^8}; \quad \gamma=W^2/M^2.$$
Here we defined $E_2(\gamma)=1-(1+\gamma+\gamma^2/2)e^{-\gamma}$ and
\begin{equation}
I_n=\int_0^{\infty}dkk^n\frac{m^2(k)}{k^2+m^2(k)}.
\label{23}
\end{equation}
In the term $\tilde A'_4(M^2, W^2)$ the gluon condensate is included. It is
\begin{equation}
\tilde A_4'=\frac{bM^2E_0(\gamma)}{4L(M^2)}-\frac{3bI_3}{2M^2}; \quad b=(2\pi)^2\langle 0|G^{a\mu \nu}G^a_{\mu \nu}|0\rangle,
\label{24}
\end{equation}
while $E_0(\gamma)=1-e^{-\gamma}$.

While the contributions $\tilde A'_0(M^2, W^2)$ and $\tilde A'_4(M^2, W^2)$, in which the integrands contain the value
$m^2(k)$ are saturated by $k \sim 1/\rho \ll q$, the term  $\tilde B'_3(M^2, W^2)$ contains the factor $m(k)$ in the integrand. The region $k \sim q$ becomes important. Here we need a parametrization of the propagator $S_I(q)$. We employ
\begin{equation}
S_I(k)=\frac{{\cal A}}{(p^2+\eta^2)^4},
\label{25}
\end{equation}
The power of denominator is determined by the asymptotic behavior
$S_I(k) \sim k^{-8}$ at $k \rightarrow \infty$ \cite{9}. The parameters ${\cal A}$ and $\eta$ are fixed by the values
of the effective quark mass $m(0)$ and by the value of the scalar condensate
\begin{equation}
\langle 0|\bar q(0)q(0)|0\rangle= i\int\frac{d^4p}{(2\pi)^4}Tr S_I(p)=-4N_c\int\frac{d^4p}{(2\pi)^4}\frac {m(p)}{p^2+m^2(p)}.
\label{26}
\end{equation}
Thus
\begin{equation}
\frac{{\cal A}}{\eta^8}=\frac{1}{m(0)}; \quad \frac{{\cal A}}{2\eta^4}=a; \quad a=-(2\pi)^2\langle 0|\bar q(0)q(0)|0\rangle \rangle0.
\label{27}
\end{equation}

Employing parametrization (\ref{25}) we find
\begin{equation}
\tilde B'_3=2aM^4F_3(M^2);
\label{28}
\end{equation}
$$ F_3(M^2)=e^{-\beta}(E_1(\gamma)-\beta E_0(\gamma) + \beta^2({\cal E}(\beta)-{\cal E}(\gamma))-\frac{2I_3}{M^4}; \quad \beta=\eta^2/M^2.$$
Here $E_1(\gamma)=1-(1+\gamma)e^{-\gamma}$, while ${\cal E}(\beta)=\int_{\beta}^{\infty}dte^{-t}/t$.
Also
\begin{equation}
A_6'=\frac{4}{3}a^2F_6(M^2); \quad
F_6(M^2)=\frac{\eta^4}{M^4}\int_0^1\frac{dtt\exp({-\eta^2/t(1-t)M^2})}{(1-t)^3},
\label{29}
\end{equation}
and
$ B_6'=\frac{2}{3}a^2\frac{\cal A}{M^6}.$

Now we write down the dependence of the input parameters on the actual values of $\rho$ and$R$. We denote by the lower index $0$ the values corresponding to $\rho_0$ and $R_0$ defined by Eq.~(\ref{4}). We introduce
\begin{equation}
\zeta=\frac{\rho}{\rho_0}; \quad \kappa=\frac{R}{R_0}.
\label{290}
\end{equation}
We find immediately that for the gluon condensate (see Eq.~(9)) $b=b_0/\kappa^4$ and also \cite{9}
\begin{equation}
a=\frac{a_0}{\zeta \kappa^2}; \quad m(0) =m_0(0)\frac{\zeta}{\kappa^2}.
\label{291}
\end{equation}
Also, since $I_3=2\pi^2/N_cR^4$, one finds $I_3=I_{30}/\kappa^4$. However, in order to obtain the dependence of the integrals $I_5$ and $I_7$ one should carry out the computations of the functions $m(p;\rho,R)$ for various values of $\rho$ and$R$.

\section {Variation of the value of the instanton size}

Now we modify the values of $\rho$ keeping $R=R_0=1~$fm. In this case $\kappa=1$. For the parameters of parametrization
given by Eq.~(\ref{25}) we find $\eta^2=\eta_0^2$ while ${\cal A}={\cal A}_0/\zeta$. Also the scalar quark condensate
$a=a_0/\zeta$. Note that the condensate $a$ is related to the normalization point $\mu =1/\rho \approx 600$ MeV. The value at the conventional point $a_c$ can be presented as $a_c=aL(\mu_c^2)$ with
\begin{equation}
L(\mu_c^2)=\Big(\frac{\ln \mu_c^2/\Lambda^2}{\ln \mu^2/\Lambda^2}\Big)^{4/9},
\label{31} \end{equation}
where $\Lambda=\Lambda_{QCD}$ is the QCD scale.
Thus $a_c \approx 0.9a$.

The  results of calculations  is presented in Table 1. The consistency of the LHS and RHS of the SR is illustrated by Fig.1. Starting with the case of independent particle approximation
(the terms $A_4$ are not included) we see that the value of the nucleon mass is very stable, changing by about several percent while $\rho$ changes by about $10\%$ around the point $\rho=\rho_0$. The same estimations are true for the case when
the correlations described by the term $A_4$ are included. The results for a larger interval of the values of $\rho$ are shown in Fig.2.

\section {Simultaneous  variation of parameters}

Here we shall modify the values of the parameters $\rho$ and $R$ simultaneously.
We start by changing  $R$ and $\rho$ keeping $\rho/R=const=1/3$. Thus $\kappa=\zeta$.

The results of computations are presented in Table 2. One can see that in the independent particle approximation
the value of the mass changes only by about 20 MeV while we change the values of $\rho$ and $R$ by about $10 \%$.
The variation of the nucleon residue and that of the nucleon size $r_N$ is more pronounced. It increases by about
$10 \%$ if we decrease the magnitudes of $\rho$ and $R$ by these values. Also the value of the effective continuum threshold diminishes, making the ''pole+ continuum" model of the spectrum less reliable. Note that the value of the scalar condensate also becomes noticeably smaller.

Note that the accuracy of the solution of the SR equations become worse while we diminish the value of $\rho$.
We illustrate this statement by demonstrating the LSH and RHS of the SR at $\zeta=0.95$. For $\zeta \leq 0.90$
the value of $\chi^2$ per point exceeds unity. Thus we would rather state that there is no solution for these
values of $\zeta$. Note also that the value $\zeta=0.95$ corresponds to  $a=0.816$ GeV$^3$, i.e. for normalization at $\mu_c=0.5$GeV $\langle 0|\bar q(0)q(0)|0\rangle=(-278$MeV$)^3$

If the correlations caused by the gluon condensate are involved the results for variation of the nucleon mass do not change much. However, the values of $\lambda^2$ and $W^2$ become more stable while we change the value of $\rho$. The results are also given in Table 2.

Now we change $\rho$ and $R$ keeping the scalar condensate $a=a_0$ to be constant and putting $a_0=0.7$ GeV$^3$. Thus $\zeta=1/\kappa^2$,
$\rho \sim R^{-2}$, and variation of $R$ provides a twice stronger variation of $\rho$. As well as in previous cases the value of $m$ increases with increasing of $\rho$. In contrast to the previous cases the values of $\lambda^2$ and $W^2$
increase with increasing of $\rho$. In all cases the $\rho$ dependence becomes more smooth if we include the gluon condensate - see Table 3. Similar tendency remains for a smaller value $a_0=0.6$ GeV$^3$. However in this case the influence of correlations appears to be much stronger, shifting the value of the nucleon mass from $1.11$ GeV (at $\rho=\rho_0$)
to $m=0.68$ GeV.

 \section{Summary}

We analyzed  the dependence of the solutions of the QCD sum rules equations in instanton medium on the average size of the instantons $\rho$ and on distance
of between the instantons $R$. We considered several versions of varying the values of $\rho$ and $R$. We
tried the version with $R=const$, $\rho/R=const$ and also kept the scalar condensate $\langle 0|\bar q(0)q(0)|0\rangle$
to be constant. In the latter case $\rho \sim 1/R^2$. We also compared the results in the independent particle approximation and those with inclusion of the correlations due to the gluon condensate, in which two gluons interact with two different quarks.

The general tendency is that the nucleon mass drops while we increase the value of $\rho$. The average modification of the value is about $50$ MeV while we modify the value of $\rho$ by $10\%$. In the cases $R=const$ and $\rho/R=const$ this change is accompanied by decreasing of the values of $\lambda^2$. and $\rho$. This corresponds to increase  of the size of the nucleon by about $6\%$. Also the continuum threshold becomes smaller, corresponding to lower masses of the excited states. For $\rho/R=const$ the larger value $\delta r_N/r_N \approx 0.09$ without correlations changes to a smaller one (about $0.02$) after the correlations are included.

Keeping the constant value of the scalar condensate we find that  the values of $\lambda^2$ and $W^2$ decrease while
$\rho$ increases. This makes the "pole+continuum" model for the RHS of the SR more reliable.
Here the size of the nucleon drops while we increase $\rho$ changing as $\delta r_N/r_N \approx -0.02$
for $\zeta =1.1$. The change becomes much smaller after we include the correlations.

We acknowledge the partial support by the RFBR grant 12-02-00158 and by the RSGSS grant 4801.2012.2.

{}

\clearpage
\newpage
\begin{table}
\caption{Behavior of the solution with variation of $\rho$ at $R=1~$fm.
The numbers above the horizontal line correspond to independent quarks in the polarization operator.
The numbers below the horizontal line correspond to approximate inclusion of correlations.}

\begin{center}
\begin{tabular}{|c|c|c|c|c|} \hline
$\zeta$&$m$, GeV& $\lambda^2$, GeV$^6$ &
$W^2$, GeV$^2$ & $a$, GeV$^3$\\
\hline
0.90&1.10&1.56&2,65&0.82 \\

1.00&1.13&1.27&2.47&0.70\\

1.10&1.18&1.08&2.36&0.60\\

\hline

0.90&0.85&1.08&2.24&0.82 \\

1.00&0.82&0.80&2.06 &0.70\\

1.10&0.79&0.62&1.94&0.60\\

\hline
\end{tabular} \end{center}
\end{table}

\begin{table}
\caption{Behavior of the solution with variation of $R$ and $\rho$ at $\rho/R=1/3$.
The numbers above the horizontal line correspond to independent quarks in the polarization operator.
The numbers below the horizontal line correspond to approximate inclusion of correlations.}

\begin{center}
\begin{tabular}{|c|c|c|c|c|} \hline
$\kappa$&$m$, GeV& $\lambda^2$, GeV$^6$ &
$W^2$, GeV$^2$ & $a$, GeV$^3$\\
\hline
0.95&1.15&1.94&3.09&0.82 \\

1.00&1.13&1.27&2.47&0.70\\

1.05&1.13&0.88&2.10 &0.60\\

\hline

0.95&0.89&1.31&2.55&0.82 \\

1.00&0.82&0.80&2.06 &0.70\\

1.05&0.79&0.55&1.79 &0.60\\

\hline
\end{tabular} \end{center}
\end{table}

\begin{table}
\caption{Behavior of the solution with variation of $R$ and $\rho$ at fixed value of the scalar condensate
$a=0.70~$~GeV$^3$ - see the text.
The numbers above the horizontal line correspond to independent quarks in the polarization operator.
The numbers below the horizontal line correspond to approximate inclusion of correlations.}

\begin{center}
\begin{tabular}{|c|c|c|c|} \hline
$\zeta$&$m$, GeV& $\lambda^2$, GeV$^6$ &
$W^2$, GeV$^2$\\
\hline
0.90&1.07&1.11 &2.22\\

1.00&1.13&1.27&2.47\\

1.10&1.19&1.42&2.76 \\

\hline
0.90&0.80&0.78 &1.92\\
1.00&0.82&0.80 &2.06\\
1.10&0.84&0.82&2.21 \\

\hline
\end{tabular} \end{center}
\end{table}

\newpage

\clearpage


\begin{figure}[ht] 
\centerline{\epsfig{file=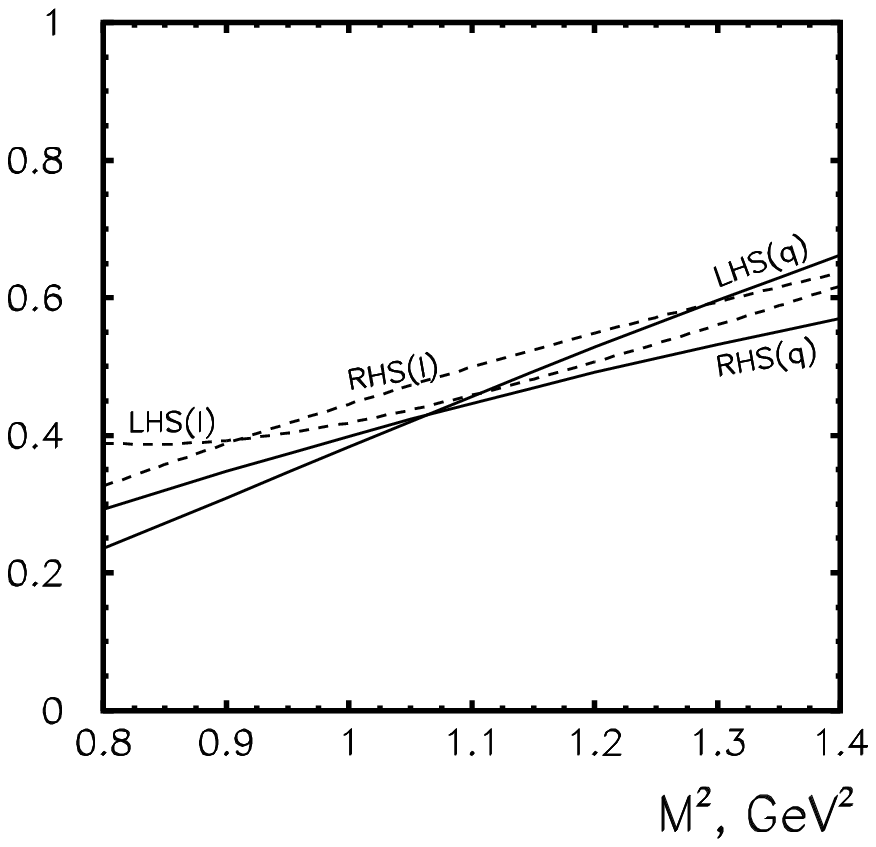,width=8.0cm}}
\caption{ Consistency of the $LHS$ and $RHS$ (given in powers of GeV) of the sum rules for the case $\rho=0.95\rho_0$, $R=R_0$.}
\end{figure}

\begin{figure}
\centerline{\epsfig{file=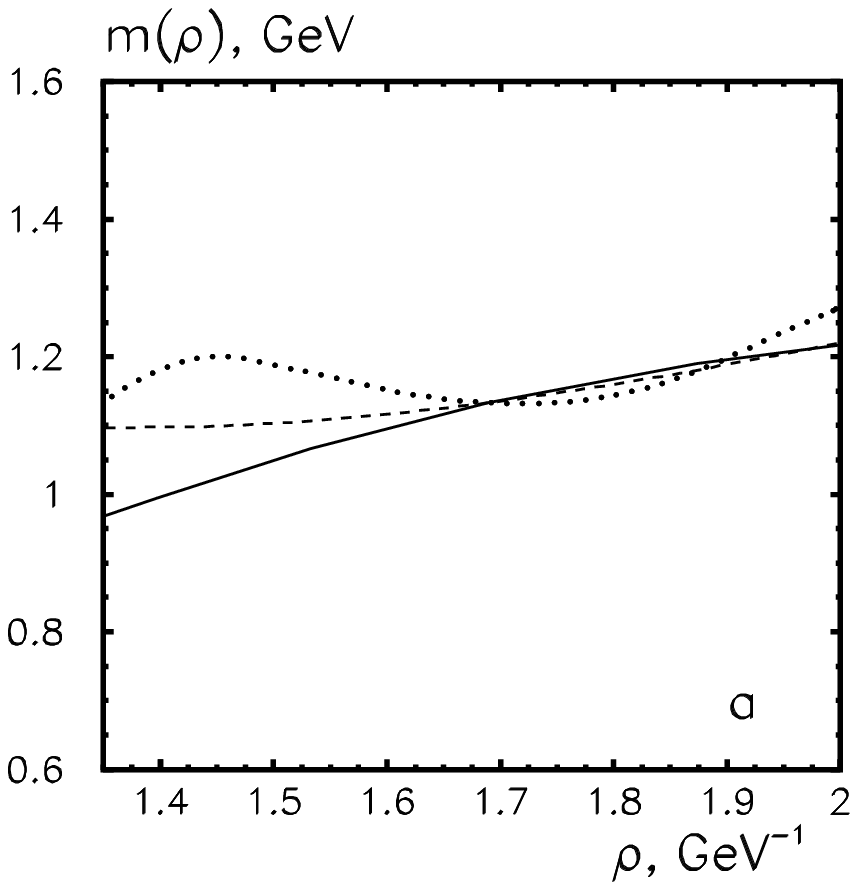,width=7cm}
\epsfig{file=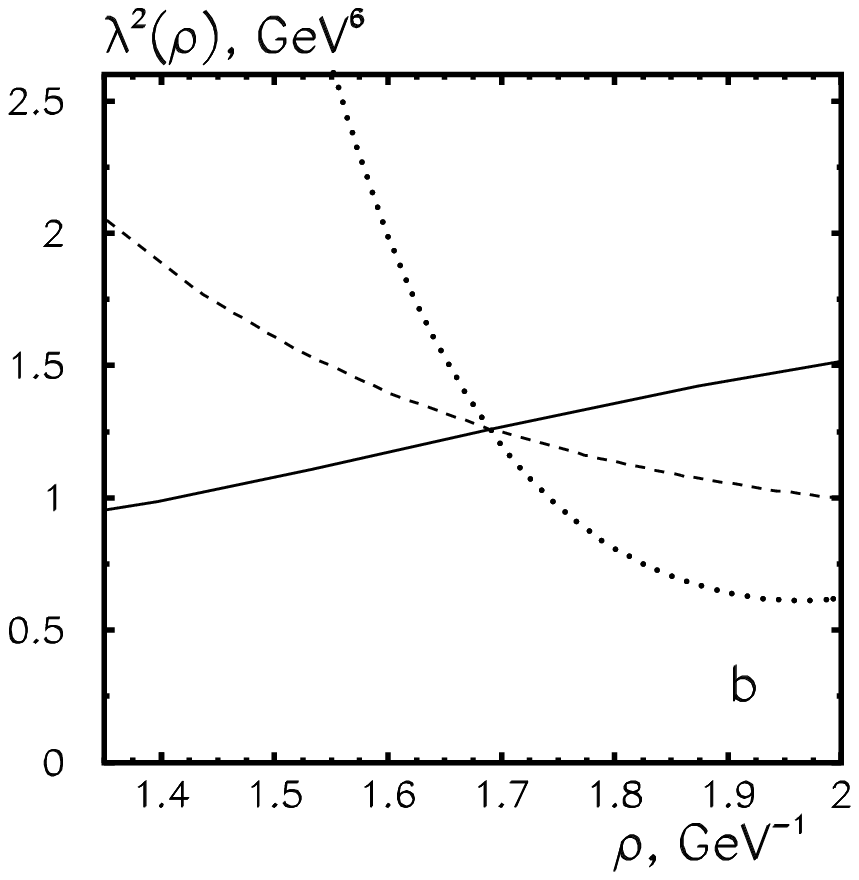,width=7cm}}
\centerline{\epsfig{file=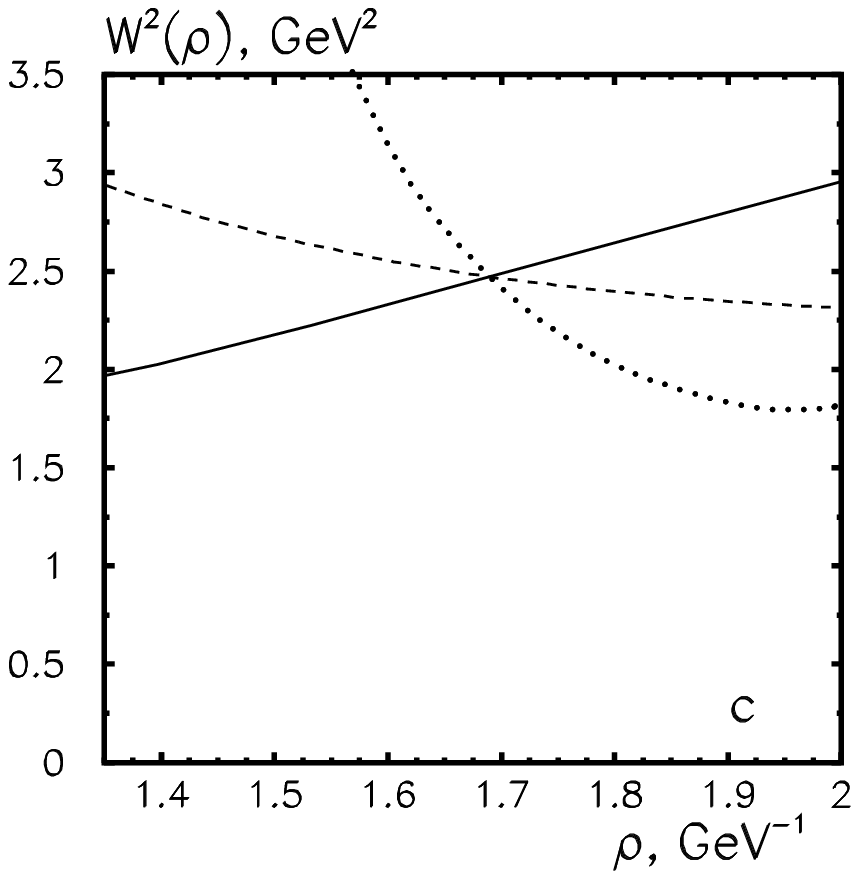,width=7cm}}
\caption{ Dependence of the solutions of the sum rules equations (given in powers of GeV) on the instanton size given in GeV$^{-1}$.
 The case $R=const=1$~fm (dashed line); $\rho/R=const=1/3$ (dotted line); $a=const=0.7$~GeV$^3$ (solid line).
Fig.~$a$ is for $m(\rho)$, Fig.~$b$ is for $\lambda^2(\rho)$, Fig.~$c$ is for $W^2(\rho)$.}
\end{figure}

\begin{figure}
\centerline{\epsfig{file=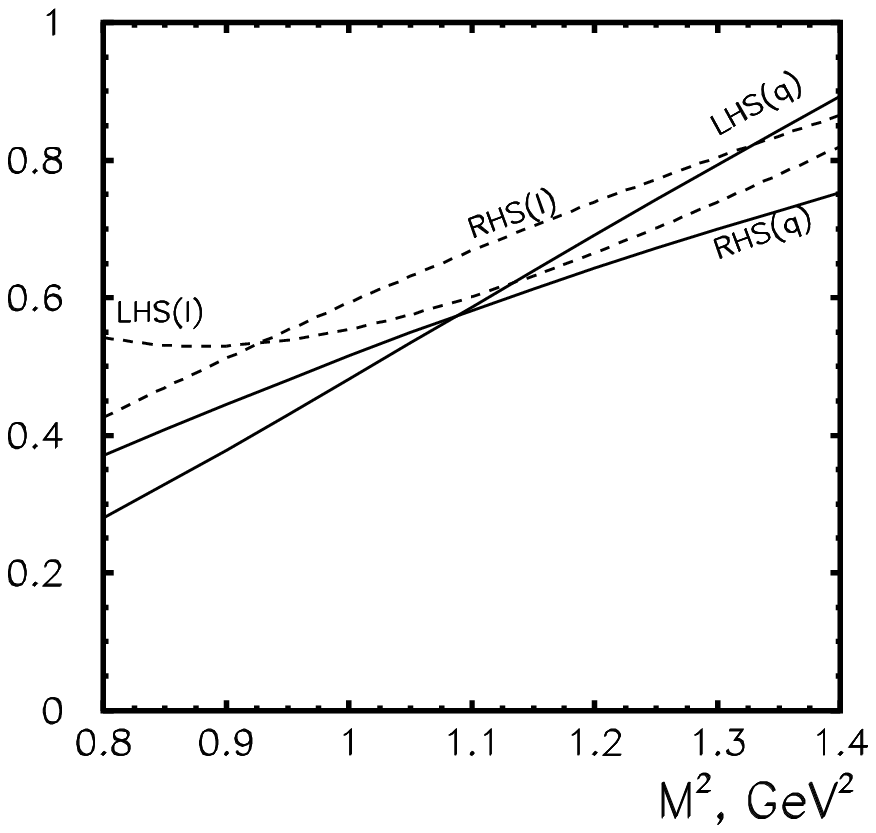,width=8cm}}
\caption{Consistency of the $LHS$ and $RHS$ (given in powers of GeV) of the sum rules for the case $\rho=0.95\rho_0$, $R=0.95R_0$.}
\end{figure}

\end{document}